# Modern Solid Electrolytes for All-Solid-State Batteries: Materials Chemistry, Structure, and Transport

Denys Butenko[1], Mustafa Khan[1], Liusuo Wu*[,2], Jinlong Zhu*[,1,2]

[1] Shenzhen Key Laboratory of Solid State Batteries & Guangdong Provincial Key Laboratory of Energy Materials for Electric Power & Guangdong-Hong Kong-Macao Joint Laboratory for Photonic-Thermal-Electrical Energy Materials and Devices & Institute of Major Scientific Facilities for New Materials & Academy for Advanced Interdisciplinary Studies, Southern University of Science and Technology, Shenzhen 518055, China

[2] Department of Physics & State Key Laboratory of Quantum Functional Materials & Guangdong Basic Research Center of Excellence for Quantum Science, Southern University of Science and Technology, Shenzhen 518055, China.

* Corresponding author: L. Wu: wuls@sustech.edu.cn; J. Zhu: zhujl@sustech.edu.cn

## Abstract

Solid-state electrolytes are central to the development of all-solid-state batteries because their structures simultaneously govern ion transport, chemical stability, interfacial behaviour and manufacturability. In this review, from crystallographic symmetry to amorphous local polyhedra arrangement and combinations, we examine inorganic solid-state electrolytes through the lens of structure–property relationships, with oxides, sulfides, and halides representing three major framework chemistries. Halide solid electrolytes and their derivatives, including mixed-anion halides and antiperovskite-related materials, have expanded this landscape further by introducing new ways to regulate local coordination chemistry, defect populations, and transport-active frameworks. Across these families, fast ion conduction depends not simply on composition or crystallographic diffusion pathways, but on the coupled effects of framework topology, site-energy distribution, defect chemistry, bottleneck response, and local anion flexibility. Oxides illustrate transport within chemically robust but geometrically constrained frameworks. Sulfides demonstrate that a soft, easily polarizable lattice can broaden the array of low-energy migration pathways. Halides occupy an intermediate state, in which the closely packed anion sublattices, an approximately degenerate lithium environment, and mixed anion coordination enable effective transport while simultaneously enhancing oxidation stability and compatibility with cathodes. Building on these comparisons, we argue that long-range ion transport is increasingly understood not as motion along a single idealized pathway, but as the macroscopic outcome of statistically connected low-barrier local migration events distributed across the structure. We further discuss the experimental and computational approaches required to establish such multiscale structure–property relationships and outline future strategies for designing transport-active frameworks in which conductivity, stability, and processability are optimized together.

## 1. Introduction

The development of all-solid-state batteries has placed solid electrolytes at the center of materials design, because their atomic-scale structures directly govern cell-level performance[1-4]. In liquid-electrolyte batteries, ionic transport is largely decoupled from long-range structural order because the electrolyte is already an ideal dynamically disordered medium. In inorganic solid-state electrolytes, by contrast, ion migration is inseparable from the architecture of the host lattice. The identity of the anion framework, the topology of available interstitial sites, the geometry of bottlenecks, the distribution of defects, and the degree of local structural flexibility together define whether alkali ions experience a deeply trapped landscape or a percolating network of accessible migration events[5-8]. For this reason, the discovery and optimization of solid-state electrolytes is fundamentally a structure–property problem rather than a simple search for the highest bulk conductivity. Looking for electrolytes that hold high ionic conductivities, stable chemical windows, as well as a large working temperature window, high safety requirements, low bulk modulus ("soft"), amorphous electrolytes become one potential candidate material. Moreover, practical electrolytes are property driven with whatever structure or even amorphous state, such as liquid + solid electrolytes, SEI, or CEI with complex composition and architecture. However, to explore these specific and complicated states, we can learn from the traditional theoretical paradigm of crystallographic symmetry, and start from fundamental polyhedral topology, combination, architecture, statistics, experimentally and theoretically investigated "structure"-property relationship.

This perspective is especially important because the performance of a solid electrolyte is never determined solely by ionic transport[9,10]. For practical implementation in all-solid-state batteries, solid electrolytes must combine sufficiently fast ion conduction with chemical compatibility with electrodes, electrochemical stability over a relevant voltage range, acceptable mechanical properties during fabrication and cycling, and processability compatible with practical cell architectures[11-13]. These requirements are difficult to satisfy simultaneously because they are often coupled through the same structural chemistry. A rigid framework may improve thermal and environmental robustness while narrowing diffusion bottlenecks. A soft, polarizable anion lattice may lower migration barriers while also increasing moisture sensitivity and interfacial reactivity. Defect-rich and locally heterogeneous structures may enhance transport by broadening the accessible energy landscape, but the same chemical complexity can also alter decomposition pathways or destabilize the framework during operation. Structure–property relationships in solid electrolytes are therefore intrinsically multidimensional: transport, stability, mechanics, and processability cannot be treated as independent axes[14-18].

Among inorganic solid-state electrolytes, oxides, sulfides, and halides represent three major structural strategies for achieving ion mobility. Oxides derive their strengths from strongly bonded oxygen frameworks, which typically provide high chemical and thermal robustness but often at the expense of lattice softness and facile processing[19-21]. Sulfides occupy the opposite end of the design space, where highly polarizable sulfur frameworks lower migration barriers and enable exceptional room-temperature conductivity, but introduce narrower practical stability windows and greater sensitivity to moisture and interfacial chemistry[22-24]. Halides have emerged more recently as a particularly attractive intermediate platform. Close-packed halogen sublattices can generate quasi-degenerate tetrahedral and octahedral Li environments, allowing fast transport through subtle local anion displacements, while often retaining better oxidative stability and cathode compatibility than sulfides[25-28]. Within the broader halide field, mixed-anion derivatives and antiperovskite-type phases have further expanded the design landscape by introducing new ways to regulate site energies, defect chemistry, framework rigidity, and dynamic transport assistance[29-32]. Conventional halides, dual-anion halides, and halide-derived antiperovskites therefore now form a particularly fertile ground for understanding how local coordination diversity, defect regulation, and structural complexity can be used to design fast-ion conductors.

Historically, the field often interpreted ion transport in crystalline solids in terms of a few idealized diffusion pathways. This language remains valuable, particularly in highly ordered frameworks where the topology of available sites can indeed be mapped onto crystallographic channels. Yet the increasing diversity of modern solid electrolytes, especially soft-lattice sulfides, mixed-anion halides, disordered halide derivatives, and dynamically active antiperovskites, has made it clear that this description is not always sufficient. In many of the most promising materials, long-range ion transport emerges not from repeated migration along a single canonical route, but from the collective connectivity of many local low-barrier hops distributed over a chemically and structurally heterogeneous landscape[25,33-35]. This shift in viewpoint is particularly timely because it allows crystalline and disordered conductors to be discussed within a common conceptual framework. Rather than asking only whether a material contains a specific diffusion pathway, one can ask whether its local structure generates a statistically connected network of energetically accessible migration events.

This review adopts that structure–property perspective. After classifying inorganic solid-state electrolytes according to their dominant anion chemistry and structural framework, and summarizing the characteristic transport mechanisms of oxides, sulfides, halides, and halide-derived antiperovskite systems, we draw these families together through a set of cross-family design principles for fast-ion transport. Particular emphasis is placed on the transition from pathway-defined transport in highly ordered crystals to connectivity-governed transport in materials with increasing local disorder, mixed-anion complexity, or dynamic anion participation. We then discuss the experimental and computational tools required to establish such structure–property relationships across multiple length and time scales. By focusing on the interplay among framework topology, local coordination diversity, defect chemistry, dynamic structural response, and transport connectivity, this Review aims to provide a unified picture of how fast-ion conduction can be designed across inorganic electrolyte families, and why halide solid electrolytes and their derivatives now occupy such a central place in the field.

## 2. Classification of inorganic SSEs

Inorganic solid-state electrolytes are commonly categorized by their dominant anion chemistry and structural framework, with oxides, sulfides, and halides representing the principal material classes in contemporary solid-state battery research. This classification reflects more than a difference in chemical composition, because each family is associated with a distinct set of structure–property relationships that govern ionic transport and electrochemical function. Differences in bonding character, coordination chemistry, polyhedral arrangement, lattice polarizability, defect formation, and diffusion-network topology result in substantial variations in ionic conductivity, activation energy, electrochemical stability, interfacial reactivity, mechanical properties, and environmental tolerance[36-39]. As a result, each class offers a characteristic combination of advantages and limitations, and no single family simultaneously maximizes all relevant performance metrics for practical all-solid-state batteries. A classification of inorganic SSEs based on chemistry and structure, therefore, provides a necessary framework for comparing material families and for understanding how fundamental structural features shape electrolyte behavior. This classification and structure–property design framework, along with the associated trade-offs among ionic conductivity, stability, and processability across different material classes, is schematically illustrated in Figure 1.

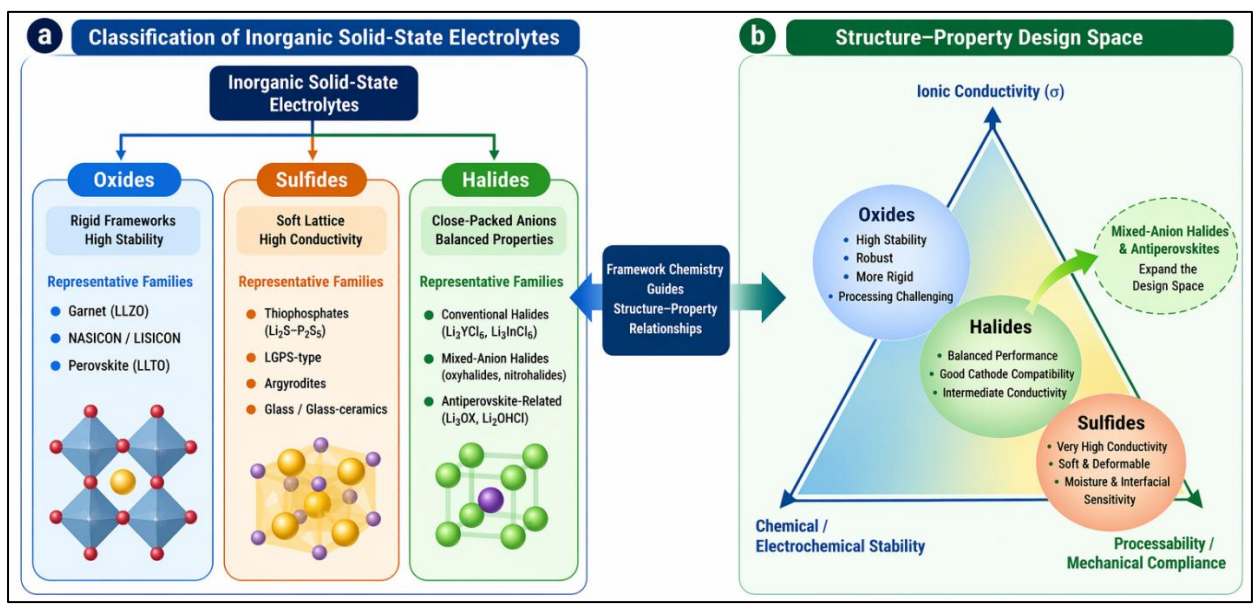

**Figure 1. Classification and design of inorganic solid-state electrolytes.** (a) Classification of inorganic solid-state electrolytes into oxides, sulfides, and halides based on anion chemistry and structural framework, with representative subfamilies. (b) Structure–property design space showing trade-offs among ionic conductivity, chemical/electrochemical stability, and processability. Oxides, sulfides, and halides occupy distinct regions, while mixed-anion halides and antiperovskite-related materials expand the accessible design space.

## 2.1 Oxides solid electrolytes

Oxide solid-state electrolytes represent one of the most established classes of inorganic ion conductors and have played a foundational role in shaping the development of solid-state batteries. Their structures are typically built from strongly bonded oxygen polyhedra, which generate rigid anion frameworks with well-defined cation sublattices and relatively stable diffusion environments. This structural feature gives oxide electrolytes their characteristic chemical and thermal robustness, but it also makes ionic transport highly sensitive to the framework's detailed geometry[40-43]. In oxide systems, $Li^+$ or $Na^+$ migration is governed not simply by composition, but by the cooperative effects of polyhedral connectivity, bottleneck size, site topology, cation ordering, and defect distribution within the oxygen framework. Subtle variations in these parameters can markedly alter the continuity of diffusion pathways, the local site-energy landscape, and the migration barrier for long-range ion transport[44-46]. As a result, oxide electrolytes often exhibit a distinctive balance of properties: they generally offer strong chemical robustness and good environmental tolerance, but their rigid lattices can impose relatively high migration barriers, while dense ceramic processing and grain-boundary resistance often become major practical constraints. These structure–property relationships in oxide electrolytes, linking rigid framework geometry, migration pathways, and microstructural limitations to ionic transport behavior, are schematically illustrated in Figure 2.

Among oxide electrolytes, garnet-type materials, especially $Li_7La_3Zr_2O_{12}$ (LLZO), have become the most influential model system in modern all-solid-state battery research. The garnet framework is constructed from a rigid three-dimensional network of corner-sharing $ZrO_6$ octahedra and La-centered polyhedra, within which lithium occupies a partially filled interstitial sublattice[41,47-49]. The key structural feature of this family is a three-dimensional Li migration network, which allows ions to move through interconnected tetrahedral and distorted octahedral/interstitial sites. In the highly conducting cubic phase, the lithium distribution is relatively disordered, and the local energy landscape is comparatively flat, facilitating Li hopping between neighboring sites and supporting higher room-temperature ionic conductivity. By contrast, the tetragonal phase exhibits a more ordered lithium arrangement, stronger site differentiation, and reduced pathway degeneracy, all of which suppress long-range ion transport and lead to much lower conductivity. In this sense, LLZO clearly demonstrates that increased site disorder and improved three-dimensional connectivity can be beneficial for ion transport in rigid oxide lattices. At the same time, garnet electrolytes also reveal the practical limitations of oxide SSEs. Although they combine relatively high bulk conductivity with low electronic conductivity and apparent compatibility with lithium metal, their performance is often compromised by difficult densification, lithium loss during high-temperature sintering, and surface contamination by $Li_2CO_3$ or LiOH, which increases interfacial resistance. More recent studies have further shown that grain-boundary chemistry plays a decisive role in determining lithium penetration and electrochemical stability, indicating that the useful performance of garnet electrolytes depends not only on bulk composition but also on microstructural and interfacial design[21,46,50]. Open-framework oxides, including NASICON- and LISICON-related electrolytes, provide another important structural route to fast ion transport. In these materials, ionic migration is enabled by interconnected interstitial pathways embedded within an oxygen-based framework, although the detailed topologies of mobile-ion sites and the structural origins of pathway connectivity differ between the two families[51,52]. In NASICON-type systems, exemplified by $LiTi_2(PO_4)_3$, LATP, and LAGP, the framework is built from corner-sharing $MO_6$ octahedra and $XO_6$ polyanions, which generate an open three-dimensional skeleton containing interconnected cavities and diffusion channels for alkali ions[52,53]. The main structural advantage of this family lies in its framework openness: when bottlenecks are sufficiently wide, and the site network remains continuous, relatively fast bulk ion transport can be achieved. In LISICON-derived oxides, by contrast, transport generally proceeds through networks of tetrahedral or mixed-coordinate sites, and the conductivity depends strongly on how these sites are connected and how evenly their site energies are distributed[54-56]. In both cases, the essential structure–property relationship is similar: ionic transport improves when the framework provides continuous migration pathways, accessible interstitial sites, and a favorable balance between site stability and pathway connectivity, whereas transport deteriorates when bottlenecks become constricted or site energies become too uneven. These systems therefore illustrate particularly well how local polyhedral geometry, framework distortion, and compositional tuning can directly reshape ion-migration barriers. At the same time, their favorable bulk transport characteristics do not fully eliminate the common limitations of oxide electrolytes. In particular, Ti-containing NASICON phases are readily reduced by lithium metal, whereas grain boundaries and insufficient ceramic densification often introduce substantial resistance in practical pellets. Thus, open-framework oxides combine structurally favorable diffusion topologies with the persistent microstructural and interfacial challenges typical of oxide SSEs.

Perovskite-related oxides, represented most prominently by lithium lanthanum titanate (LLTO), offer another classic example of how favorable crystallographic transport pathways do not necessarily translate into practical electrolyte performance[40,57-59]. In LLTO, the framework is formed by corner-sharing $TiO_6$ octahedra, whereas lithium migrates through the partially occupied A-site sublattice, where La, Li, and vacancies together define the available diffusion channels. In this family, A-site deficiency is a critical structural parameter because it creates the vacancies required for lithium transport and strongly influences channel connectivity. When the vacancy arrangement and local bottleneck geometry are favorable, LLTO can exhibit very high bulk lithium-ion conductivity, making it one of the most instructive oxide systems for studying rapid ion transport. However, this structural advantage is offset by serious practical drawbacks. Strong grain-boundary blocking often reduces the total conductivity of polycrystalline ceramics far below the intrinsic bulk value, while the presence of Ti also introduces reduction instability against lithium metal. As a result, LLTO illustrates a central lesson in oxide electrolyte design: a framework may provide highly efficient local or bulk migration pathways, yet the overall usefulness of the material can still be limited by grain-boundary resistance and chemical incompatibility at the electrode interface.

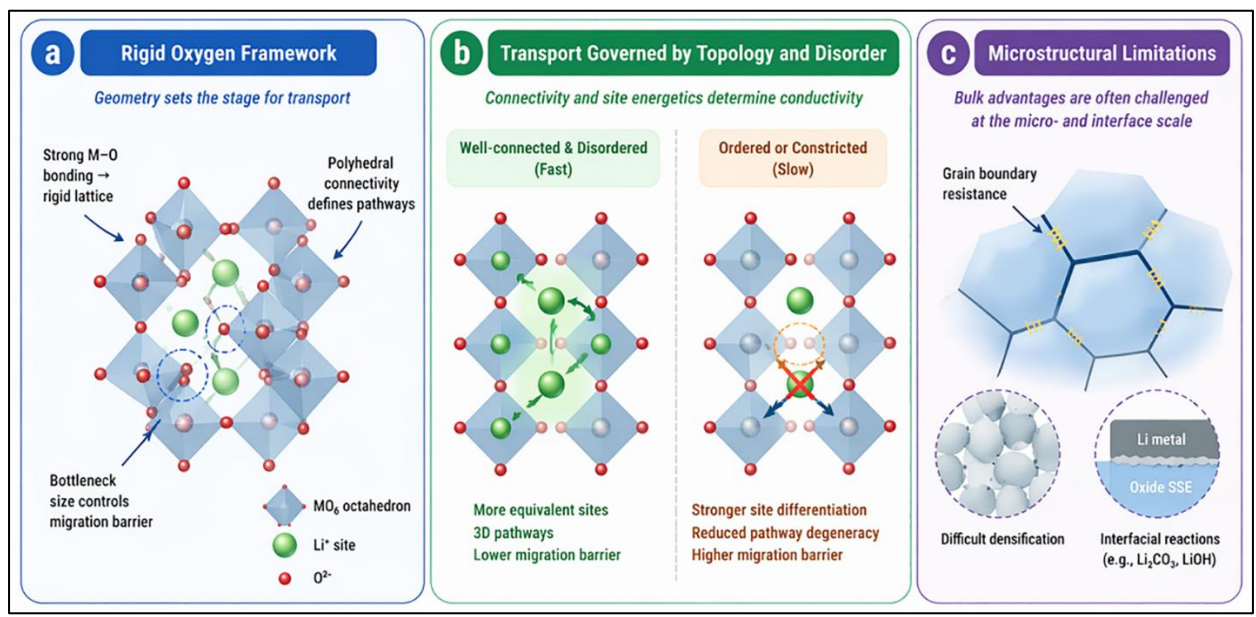

**Figure 2. Structure–property relationships in oxide solid electrolytes**. (a) Rigid oxygen frameworks define polyhedral connectivity and bottleneck-controlled ion migration. (b) Transport depends on site connectivity and disorder, with well-connected structures enabling lower migration barriers than ordered frameworks. (c) Microstructural and interfacial limitations, including grain-boundary resistance and densification challenges, often restrict practical performance.

Overall, oxide electrolytes derive their principal advantages from strong M–O bonding, which provides a robust structural and chemical foundation. However, the same framework rigidity often limits lattice flexibility and constrains ion migration compared with more polarizable anion systems. In this context, sulfide electrolytes present a distinctly different materials strategy: rather than prioritizing structural robustness and environmental tolerance, they are characterized by softer lattices, lower migration barriers, and much higher room-temperature ionic conductivities. The sulfide family therefore provides an important complementary route to fast ion conduction in solid-state batteries.

## 2.2 Sulfide solid electrolytes

Sulfide solid-state electrolytes constitute one of the most important classes of inorganic ion conductors because they combine exceptionally high room-temperature ionic conductivity with good deformability and relatively facile densification. Structurally, this family is broad rather than monolithic, encompassing glassy and glass-ceramic thiophosphates in the $Li_2S$–$P_2S_5$ system, thio-LISICON phases, LGPS-type conductors, argyrodites such as $Li_6PS_5X$ (X = Cl, Br, I), and a growing range of substituted phosphosulfides and mixed-anion derivatives[60]. Across these materials, fast ion transport is commonly associated with deformable sulfur-based frameworks, thiophosphate or related tetrahedral building units, and partially occupied alkali-ion sublattices that provide multiple accessible migration pathways[61,62]. As a result, the transport behavior of sulfide electrolytes is governed by the connectivity of the anion framework, the degree of alkali-site disorder, the dimensionality of the diffusion network, and the extent to which the local structure can adapt during ion migration. These features make sulfides the benchmark family for fast bulk ion transport, while also introducing important challenges in electrochemical stability, moisture tolerance, and interface control. These structure–property relationships in sulfide electrolytes, highlighting the role of soft polarizable frameworks, disorder-enhanced connectivity, and the resulting trade-off between fast ion transport and stability limitations, are schematically illustrated in Figure 3.

Among sulfide electrolytes, $Li_{10}GeP_2S_{12}$ (LGPS) remains the most emblematic system because it established that an inorganic solid electrolyte can achieve room-temperature ionic conductivity on the order of $10^{-2}$ S·cm$^{-1}$, approaching the transport regime traditionally associated with liquid electrolytes. Structurally, LGPS and related thio-LISICON phases are built from tetrahedral thiophosphate- and Ge-containing units, principally $(PS_4)^{3-}$ and $(PS_3)^{4-}$, which together generate a highly favorable framework for lithium migration[63]. In this structure, transport has often been described in terms of rapid motion along the c-axis, but the migration network is better viewed as multidimensional because the framework also permits diagonal hops within the ab plane and cross-linking transitions between $LiS_4$ and $LiS_6$ environments. This combination of connected pathways and locally adaptable sulfur coordination lowers the energetic cost of ion migration and produces exceptionally high bulk conductivity. LGPS therefore illustrates a central sulfide design principle: when a soft tetrahedral framework provides well-connected pathways and weakly confining site energetics, ultrafast ion transport can be achieved. At the same time, the chemistry that supports outstanding conductivity also limits practical stability, and LGPS remains highly sensitive to moisture and unstable in contact with lithium metal without deliberate interfacial protection.

A broader, chemically versatile platform is provided by thiophosphate-based Li/Na–P–S systems, in which the anion framework's structural identity changes with composition[64-66]. In these materials, the framework is constructed from thiophosphate polyanions such as orthothiophosphate $(PS_4)^{3-}$, pyrodithiophosphate $(P_2S_7)^{4-}$, and related sulfur-rich tetrahedral units. Their relative abundance depends strongly on the Li/S ratio and, more generally, on the degree of thiophosphate polymerization. Compositions rich in lithium and sulfur tend to favor frameworks dominated by isolated $(PS_4)^{3-}$ tetrahedra, which create more open and deformable local environments for alkali-ion migration. By contrast, compositions with lower Li and S contents more readily stabilize polymerized thiophosphate units, in which corner- or edge-sharing tetrahedra generate a more rigid anionic backbone. This structural distinction has direct consequences for ion transport. Frameworks rich in isolated thiophosphate units generally provide smoother and more accessible migration pathways, whereas increased polyanion polymerization tends to reduce local flexibility and alter the continuity or dimensionality of the diffusion network. In this sense, the degree of thiophosphate polymerization serves as a useful structural descriptor linking composition to transport behavior[16,23,24]. These systems therefore demonstrate that, in sulfide electrolytes, high conductivity is not determined simply by the presence of sulfur, but by how the sulfur-based polyanion framework organizes local flexibility, pathway connectivity, and alkali-site accessibility.

Argyrodite sulfides constitute another major structural class and have become especially important because they combine high conductivity, compositional flexibility, and comparatively accessible synthesis.

Representative compositions such as $Li_6PS_5Cl$, $Li_6PS_5Br$, and $Li_6PS_5I$ are built from a tetrahedrally coordinated thiophosphate framework containing sulfide and halide anion sublattices, within which lithium occupies a partially disordered network of interstitial sites[67-69]. The key structure–property feature of argyrodites is the strong coupling between anion disorder and lithium transport. Halide/sulfide site inversion, aliovalent substitution, and the associated redistribution of local vacancies can flatten the migration-energy landscape and improve connectivity among Li sites, thereby enhancing conductivity. This is why argyrodites have become model systems for understanding how controlled disorder promotes rapid ion transport in soft-lattice conductors. In optimized compositions, room-temperature ionic conductivities can approach the lower end of the $10^{-2}$ S cm$^{-1}$ range, while more typical values remain in the $10^{-3}$ S cm$^{-1}$ regime. However, the attractive transport properties of argyrodites do not eliminate the usual sulfide stability challenges. Their mechanical softness and good interparticle contact can reduce initial interfacial resistance, but they are still not intrinsically stable against lithium metal, and oxidation against high-voltage cathodes remains a major limitation. Thus, argyrodites clearly show that disorder-enhanced transport can yield excellent conductivity, while practical deployment still depends on careful control of interface chemistry.

Glassy and glass-ceramic sulfides provide a complementary perspective, showing that high ion transport in sulfide systems does not rely solely on highly ordered crystalline frameworks. In $Li_2S$–$P_2S_5$-based glasses and glass-ceramics, alkali ions migrate through a structurally heterogeneous but mechanically compliant thiophosphate network in which short-range connectivity and local dynamical flexibility remain sufficient to support rapid hopping[70-72]. These materials are important not only historically but also practically, as they can often be produced by mechanochemical synthesis, melt-quenching, or related low-temperature routes, without the extreme sintering burden characteristic of oxide ceramics. Their powder processability and mechanical compliance make them especially attractive for composite cathodes and pressure-assisted cell assembly. At the same time, their transport properties remain highly sensitive to synthesis history, local structural rearrangement, and surface chemistry, indicating that processability and electrochemical behavior are intimately coupled in these systems. Glassy and glass-ceramic sulfides, therefore, emphasize an important feature of the sulfide family as a whole: the same soft-lattice chemistry that enables rapid ion transport also enables comparatively facile processing, although often with increased sensitivity to atmosphere and interfacial degradation[73,74].

Despite their outstanding transport properties, the practical performance of sulfide electrolytes is constrained by limited electrochemical and environmental stability. Many sulfide electrolytes are now understood to possess relatively narrow intrinsic stability windows and are better regarded as metastable than truly inert under battery operating conditions. Their successful operation often depends on whether the initial decomposition products remain sufficiently ionically conductive, electronically insulating, or kinetically self-limiting to prevent continued degradation[15,75-78]. This issue is especially important at the lithium-metal interface, where reductive decomposition can generate mixed-conducting interphases, promote current localization, and accelerate impedance growth or filament penetration. Moisture sensitivity is another defining challenge of the sulfide family. Exposure to humid air can induce hydrolysis, formation of surface degradation products, conductivity loss, and, in many cases, $H_2S$ evolution. As a result, sulfide electrolytes impose a greater dry-room or glovebox burden than oxide electrolytes and are increasingly evaluated not only in terms of ionic conductivity, but also in terms of air stability and manufacturability.

These limitations have driven a shift in current sulfide research from conductivity optimization alone toward broader stabilization strategies involving compositional modification, mixed-anion design, cathode coatings, artificial interlayers, protective surface treatments, and tightly controlled processing conditions. Taken together, sulfide solid-state electrolytes remain the benchmark family for fast bulk alkali-ion transport because their sulfur-based frameworks provide a uniquely favorable environment for low-barrier ion migration. Structures built from deformable thiophosphate units, partially disordered alkali-ion sublattices, and well-connected interstitial pathways generally deliver the highest conductivities, whereas increased framework rigidity, excessive polyanion polymerization, or poorly controlled interfacial chemistry tend to compromise practical performance. At the same time, the same structural softness that enables ultrafast transport also gives sulfides their defining weaknesses in moisture tolerance, electrochemical stability, and long-term interfacial durability. Sulfide electrolytes therefore exemplify a distinct structure–property strategy in solid-state batteries: they maximize ion mobility through anion softness and framework deformability, but their successful deployment depends on whether these transport advantages can be matched by equally deliberate control of synthesis, interfaces, and operating environment.

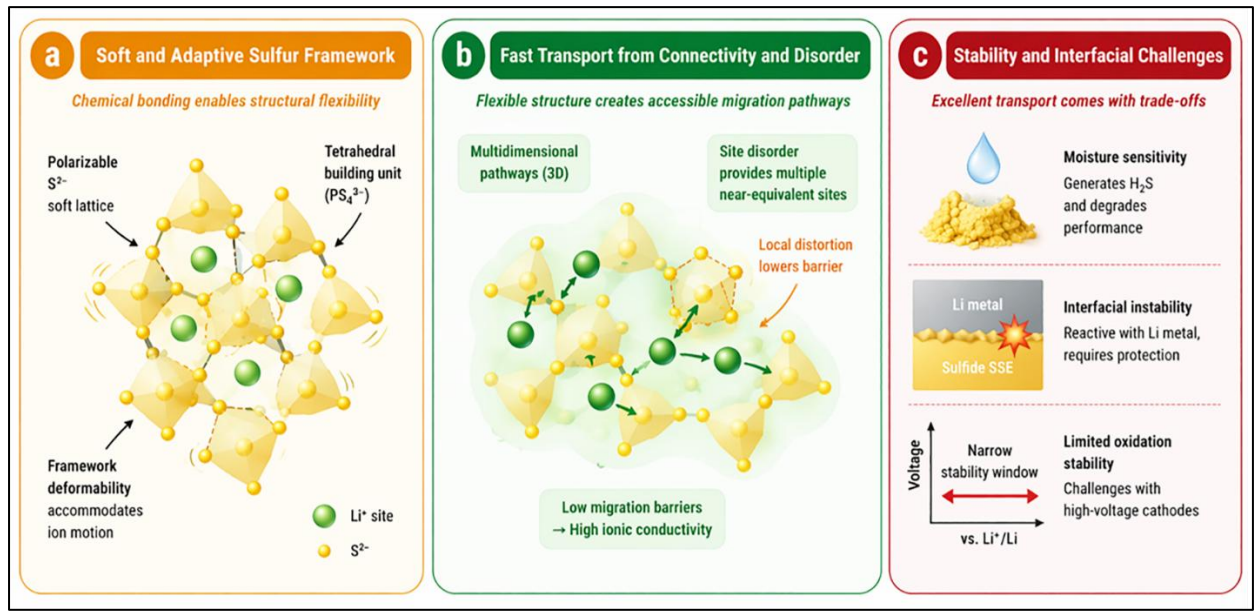

**Figure 3. Structure–property relationships in sulfide solid electrolytes.** (a) Soft, polarizable sulfur frameworks provide flexible coordination environments for ion transport. (b) Fast transport emerges from multidimensional pathways, low migration barriers, and disorder-enhanced connectivity. (c) Stability limitations, including moisture sensitivity, interfacial reactions, and narrow electrochemical windows, constrain practical performance.

## 2.3 Halide-based solid electrolytes

Halide solid-state electrolytes have emerged as one of the most dynamic classes of inorganic ion conductors, occupying an attractive middle ground in the solid-electrolyte design space. Compared with sulfides, they generally offer better oxidative stability and improved compatibility with layered oxide cathodes, while compared with oxides, they often retain softer lattices, lower-temperature processability, and more favorable mechanical compliance. For this reason, halides have become especially important in cathode-facing electrolyte design, where moderate deformability, high-voltage tolerance, and useful room-temperature ionic transport must be balanced simultaneously[39,79,80]. More fundamentally, the halide family reveals a distinct structural strategy for fast ion transport, one that does not rely on the highly rigid oxygen frameworks typical of oxides or the strongly deformable sulfur-based networks characteristic of sulfides. Instead, many halides achieve rapid migration through close-packed halogen sublattices that generate multiple weakly differentiated interstitial environments, allowing Li$^+$ transport to proceed through subtle local anion adjustments and shallow migration landscapes rather than through large-amplitude framework rearrangement. In such systems, ion migration can proceed through subtle, low-energy halide displacements without requiring large-amplitude framework rearrangement, which gives halides a combination of transport efficiency and structural coherence that is especially attractive for practical all-solid-state batteries. The evolution of halide solid electrolytes from conventional close-packed frameworks to mixed-anion systems and antiperovskite derivatives, together with their distinct structure–property relationships and transport mechanisms, is schematically illustrated in Figure 4.

Although halides are sometimes regarded as a recent breakthrough, their development has much deeper roots. During the 1980s and 1990s, simple lithium halides such as LiF, LiCl, LiBr, and LiI were already recognized as ion-conducting solids in the early twentieth century, and some were incorporated into early solid-state electrochemical devices[81-83]. However, these early materials generally suffered from very low room-temperature ionic conductivity and strong interfacial polarization, which prevented them from competing with later oxide and sulfide electrolytes. More recently (post-2018), a broader array of halide solid-state electrolytes (SSEs), encompassing spinel-type chlorides and bromides as well as trivalent-metal halides, such as Li$_3$InBr$_6$, were explored[84-87]. These investigations contributed to the field's structural chemistry; however, most halides remained less appealing than oxides or the rapidly evolving sulfides due to their generally low room-temperature conductivity or reliance on metastable high-temperature phases. A more meaningful transition occurred with halide-containing antiperovskites such as Li$_3$OCl and Li$_3$OBr, demonstrating that lithium transport in halogen-containing solids can be actively engineered through symmetry, defect chemistry, and local structural regulation[88-90]. The decisive modern breakthrough, however, came with close-packed trivalent-metal chlorides and bromides, particularly after the report of Li$_3$YCl$_6$ and Li$_3$YBr$_6$, which transformed halides from a historically interesting but underperforming class into a central modern platform for solid-state battery research. Since then, the field has expanded rapidly from simple chlorides and bromides toward mixed-anion halides, disordered halide-derived conductors and antiperovskite-related derivatives, reflecting a broader shift from searching for conductive halides to deliberately engineering halide-centered local environments for targeted transport and stability.

The modern halide field is dominated first by crystalline chloride- and bromide-based frameworks built from close-packed halogen sublattices and metal–halide octahedra. Representative compositions such as $Li_3YCl_6$, $Li_3InCl_6$, and $Li_2ZrCl_6$ derive their transport properties from halide packings that generate networks of tetrahedral and octahedral interstitial sites with small energy differences[80,85-87,91-94]. In these materials, the $MCl_6$ framework polyhedra are structurally robust, while the lithium sublattice occupies multiple quasi-degenerate coordination environments. This combination produces a relatively flat migration-energy landscape, so $Li^+$ can move through closely connected interstitial positions without being strongly trapped in any single local geometry. $Li_2ZrCl_6$ is particularly illustrative: diffraction and PDF analyses indicate that Li coordination fluctuates continuously between tetrahedral and octahedral environments, with a broad distribution of Li–Cl distances. This behavior reflects a locally adaptable halide environment in which $Cl^-$ ions undergo only subtle positional adjustments that stabilize transition states while preserving the integrity of the $ZrCl_6$ framework. The resulting transport mechanism differs from that of both oxides and sulfides. In oxides, rigid polyhedral units often impose sharper geometric constraints on hopping; in sulfides, large-amplitude distortions of soft sulfur environments often dominate the migration process. Halides instead occupy an intermediate regime in which modest anion displacements are sufficient to connect shallow energy basins through low-barrier pathways. This is a central reason why many crystalline halides can achieve high room-temperature conductivity while retaining better oxidative stability and cathode compatibility than sulfides.

Within this conventional halide branch, the structure–property relationship is especially sensitive to polyhedral connectivity and local site degeneracy. Most high-performing halides are based on cubic close-packed or hexagonal close-packed halogen lattices, but the local arrangement of metal–halide octahedra can vary considerably, with isolated, edge-sharing, or otherwise topologically connected motifs. These differences reshape the lithium migration network, alter the mechanical stiffness of the framework, and modify the distribution of accessible interstitial positions. More broadly, halide-based $Li_3MCl_6$ families illustrate that no single halide structure type is definitive; rather, the value of this family lies in how clearly it shows that ionic mobility depends on the relation between close-packed anion topology and the energetic accessibility of neighboring Li sites. When the framework preserves robust metal–halide coordination while maintaining multiple low-energy Li environments, fast transport can be achieved without sacrificing too much structural stability. However, conventional single-anion halides also exhibit a familiar limitation: it is difficult to optimize conductivity, moisture tolerance, reduction stability, and mechanical robustness simultaneously within a chemically simple chloride- or bromide-only framework. This realization is one of the main reasons the field has increasingly shifted toward dual-anion and multi-anion halide design.

Mixed-anion halides, including oxyhalides, nitrohalides, and other chemically hybrid halide-derived frameworks, have therefore become one of the most important directions in contemporary halide electrolyte research[29,95-97]. Their significance lies not merely in adding compositional complexity but in introducing a fundamentally new route to tune local coordination, defect chemistry, framework rigidity, and the topology of the lithium migration network. In conventional close-packed halides, transport is largely controlled by the halogen sublattice and the distribution of tetrahedral and octahedral interstitial sites. Once a second anion is introduced, migration is no longer governed solely by halide packing and vacancy distribution. Instead, the framework becomes chemically heterogeneous: the local coordination polyhedra around multivalent cations are modified, the relative energies of neighboring Li sites are redistributed, and the bottleneck environments along diffusion pathways become more diverse. In this sense, dual-anion chemistry should not be viewed as a minor extension of halide design, but as a distinct chemical branch in which transport, stability, and environmental response are jointly controlled by local anion complexity.

Among the various mixed-anion strategies, oxygen-containing halides are the most developed and conceptually the clearest. Early crystalline oxyhalides such as $LiNbOCl_4$ and $LiTaOCl_4$ demonstrated that introducing oxygen into a halide-derived framework does not necessarily suppress ionic transport[97-101]. On the contrary, under suitable structural conditions, oxygen can reshape the local bonding hierarchy and redistribute Li-site energies in ways that preserve or even enhance fast Li-ion migration. From a structural perspective, partial replacement of halide coordination by $O^{2-}$ modifies the cation-centered polyhedra, producing more asymmetric local environments and altering both bond polarity and bond strength. These changes can impose internal chemical pressure, distort the original metal–halide coordination shell, and flatten the energetic contrast between adjacent Li sites. As a result, oxyhalides combine part of the low-barrier migration behavior of halides with the greater local stiffness introduced by oxide ligands. This makes them especially attractive when the goal is to improve chemical robustness without fully sacrificing the transport advantages of halide frameworks. More broadly, oxyhalides show that the "best" halide design is not necessarily the softest one; in many cases, controlled local stiffening can be beneficial if it reorganizes the migration landscape into a more connected and energetically permissive network.

Nitrogen-containing mixed-anion halides extend this concept further by introducing an even stronger local perturbation to the parent halide framework. In contrast to oxygen, which often modifies the halide-derived coordination environment while still contributing an oxide-like bonding, $N^{3-}$ can drive deeper reorganization due to its higher charge density and stronger directional bonding[102,103]. When nitrogen is incorporated into Li–Zr–N–Cl or related systems, the original metal–chloride units are no longer the sole structural building blocks. Instead, mixed metal–N/Cl polyhedra and N-bridged coordination motifs emerge, broadening the distribution of local geometries and creating a more topologically diverse migration network. This is important from a transport standpoint because Li no longer migrates only through a repeated crystalline halide motif; rather, it moves through a broader population of mixed-anion environments with different local polarizabilities and site energies. In this way, nitrohalides demonstrate that a second hard anion can not only perturb the halide framework but also fundamentally reorganize it into a chemically heterogeneous conduction network. Such chemistry expands the design space well beyond traditional halides and is particularly relevant for the development of disordered or partially amorphous fast-ion conductors based on halide-derived coordination chemistry.

The mixed-anion concept also extends beyond simple oxyhalides and nitrohalides into more chemically hybridized frameworks in which oxygen-, sulfur-, nitrogen-, or polyanion-containing units cooperate to simultaneously regulate transport and stability. This broader development is important because it reveals a new design philosophy for halide electrolytes. Rather than treating the halide lattice as a fixed host that is only lightly modified by substitution, recent work increasingly treats the local anion ensemble itself as the main design variable. Additional anions can reinforce the framework, soften the local environment, diversify polyhedral distortions, reshape defect accommodation and alter the chemistry of the eventual interphase. This is particularly significant in disordered or amorphous halide-derived materials, where mixed-anion coordination broadens the range of local environments while still preserving a percolating ion-transport network. In such cases, long-range periodicity is no longer the sole determinant of fast ion migration. What matters instead is whether the local mixed-anion polyhedra remain sufficiently connected and whether the resulting energy landscape avoids deep trapping sites. This is why multi-anion halide design is becoming increasingly central to the field: it offers a route to tune ionic conductivity, moisture tolerance, oxidative stability, and processability together, rather than treating them as independent optimization targets.

Halide-containing antiperovskite materials form a structurally distinct but highly relevant branch of this broader halide landscape. They should not be viewed merely as historical precursors or peripheral analogues, but as systems that reveal a different transport topology built around halide-containing chemistry. In Li/Na-rich antiperovskites with the general formula $X_3BA$, such as $Li_2OHCl$, $Li_3OCl$, $Li_3OCl_{1-x}Br_x$, and $Na_3OCl$, the structure can be described as a three-dimensional framework of corner-sharing $OLi_6$ or $OHLi_6$ octahedra, with the halide occupying large cavities between them[32,88,104]. This electrically inverted perovskite topology places lithium on a densely packed, nearly simple-cubic sublattice that is directly connected in all three dimensions. The transport logic is therefore fundamentally different from that of conventional interstitial conductors. In ideal $Li_3OX$, the Li positions are crystallographically equivalent, so long-range transport is governed primarily by defect chemistry rather than by pre-existing site-energy asymmetry. Schottky-type LiX vacancy pairs, off-stoichiometric compositions, and aliovalent substitution, therefore, become the key variables controlling conductivity, because they determine how many mobile vacancies are available within this highly connected Li-rich framework. Antiperovskites therefore illustrate a distinct structure–property strategy within the broader halide-derived space: rather than achieving fast transport mainly by flattening the interstitial energy landscape of a close-packed halide framework, they rely on a topologically simple and intrinsically percolating Li network whose mobility is activated by defect engineering.

Within the antiperovskite family, the role of local structure and dynamic anion behavior becomes even more evident. In lithium halide hydroxides such as $Li_2OHCl$, only two-thirds of the Li vertices are occupied, leaving an intrinsic vacancy network that already supports three-dimensional percolation[105]. Neutron and AIMD studies indicate that reorientation of $OH^-$ groups within the $OHLi_6$ cages couples directly to this vacancy lattice and effectively gates Li jumps[106]. As a result, the orthorhombic-to-cubic transition and enhanced OH reorientation at elevated temperature increase conductivity by roughly two orders of magnitude. In oxyhalide antiperovskites such as $Li_3OCl$ and $Li_3OCl_{1-x}Br_x$, halogen mixing can broaden Li–X–Li bottlenecks while preserving the Li-rich framework, leading to room-temperature conductivities up to $\sim 2 \times 10^{-3}$ S·cm$^{-1}$ and activation energies as low as 0.18 eV[89,107]. Cluster-based antiperovskites such as $Li_3O(BH_4)_{0.5}Cl_{0.5}$ and $Na_3OBH_4$ introduce yet another transport mechanism: rotationally mobile $BH_4^-$ groups modulate the local electrostatic potential along migration pathways, producing a "rotating-gate" effect that can lower barriers and push ionic conductivity into the $10^{-3} - 10^{-2}$ S cm$^{-1}$ regime. These examples demonstrate that antiperovskites are not important solely because they contain halides[108,109]. They are significant because they show how symmetry regulation, defect density, bottleneck geometry, and anion dynamics can work together within a Li-rich structure to enable competitive transport while maintaining favorable reductive stability against lithium metal.

Overall, the evolution of halide solid electrolytes and their derivatives reflects a broader transition in solid-electrolyte design: from transport governed mainly by structurally prescribed interstitial pathways toward transport governed by the connectivity of low-barrier local migration environments. By combining close-packed halide frameworks, mixed-anion coordination, defect-regulated Li sublattices, and, in some cases, dynamically active local units, these materials show that fast ion conduction can be preserved across a much wider range of structural complexity than previously assumed. Halides are therefore significant not only as a high-performing electrolyte family, but also as a conceptual bridge between ordered crystalline conductors and more locally heterogeneous transport-active frameworks.

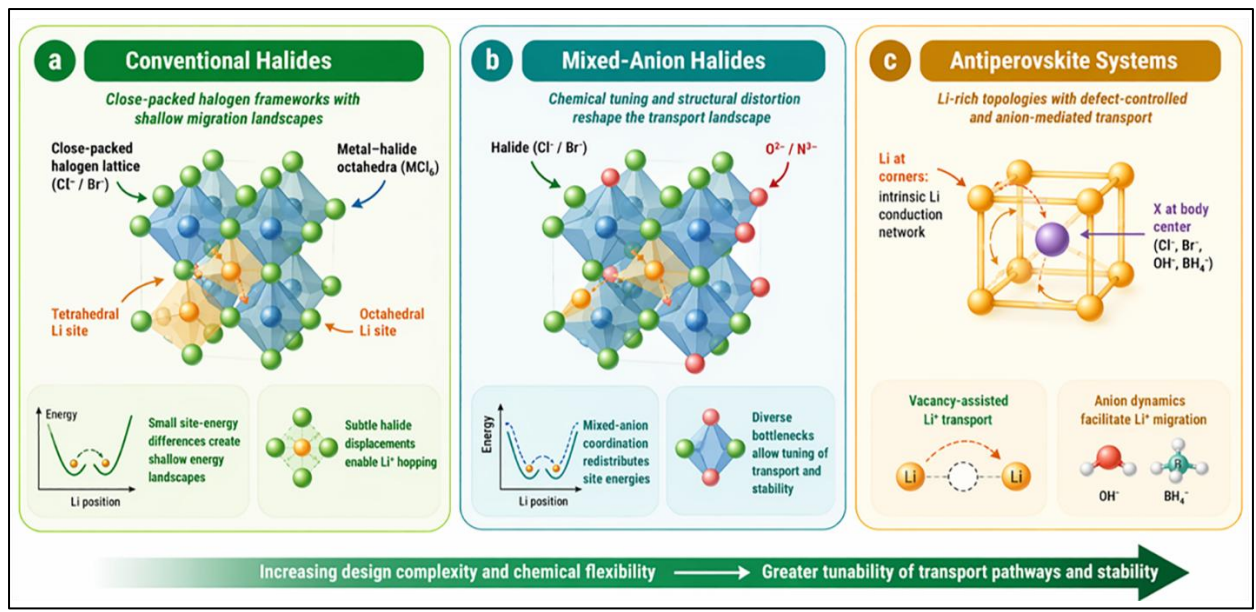

**Figure 4. Halide solid electrolytes and their derivatives: from conventional halides to mixed-anion and antiperovskite systems.** (a) Close-packed halide frameworks enable Li-ion transport through shallow energy landscapes and subtle anion displacements. (b) Mixed-anion halides redistribute local coordination and site energies, allowing tunable transport and stability. (c) Antiperovskites exhibit Li-rich, defect-controlled transport within three-dimensional percolating networks. This progression illustrates increasing chemical and structural complexity and the resulting tunability of transport and stability.

## 3. Cross-family design principles for fast ion transport

A central theme emerging across solid-state electrolyte families is that fast ion transport can no longer be understood solely in terms of a single crystallographically defined migration pathway. Such a description remains useful for highly ordered frameworks, where long-range diffusion can often be projected onto a limited set of symmetry-allowed hops between well-defined sites. However, this picture becomes progressively less sufficient as local disorder, mixed-anion complexity, polyhedral diversification, and amorphization increase. Under these conditions, long-range ion conduction is better viewed as the macroscopic outcome of statistically connected low-barrier local migration events distributed across the structure. In this broader framework, the key requirement is not simply the existence of a nominal diffusion channel, but rather whether the host lattice generates a sufficiently dense, spatially connected population of energetically accessible local environments through which ions can move without encountering deep traps or severe bottlenecks[2,6,7]. This conceptual transition from pathway-defined migration to transport governed by statistically connected low-barrier local events is schematically represented in Figure 5.

This perspective provides a useful framework for unifying transport across oxide, sulfide, halide, and antiperovskite solid electrolytes. In oxides, ion migration is often strongly constrained by rigid oxygen frameworks, and transport can frequently be related to relatively well-defined bottlenecks and crystallographic pathways. Yet even in these systems, high conductivity is not determined solely by topology. It also depends on whether vacancies, site disorder, and local distortions reduce the energetic contrast between neighboring positions sufficiently to connect otherwise restricted migration events. In sulfides, the same underlying logic appears in a more expanded form[36,75,78]. The soft and polarizable sulfur sublattice lowers the cost of local structural relaxation and broadens the range of accessible low-barrier hops, so transport is often less confined to one sharply prescribed route and more dependent on the connectivity of multiple near-degenerate migration events. Halides occupy a distinctive intermediate regime[25,80,110]. Their close-packed halogen sublattices do not usually undergo the large-amplitude distortions characteristic of sulfides, but they can still support efficient migration because tetrahedral and octahedral Li environments often differ only weakly in energy and are linked through subtle local halide displacements. Antiperovskite derivatives reveal yet another realization of the same principle: here, the Li-rich framework is already densely connected in topological terms, and the decisive issue is whether defects and dynamic anion behavior activate that latent connectivity into a transport-active network[88,104].

The decisive structural variable is therefore not the number of available sites in the abstract, but how those sites are organized into an energetically and geometrically connected migration landscape. A material may contain many interstitial positions and yet remain a poor conductor if neighboring sites are too strongly differentiated in energy or too weakly connected in space. Conversely, high conductivity can emerge when a framework creates a dense population of local environments that are both energetically similar and linked by bottlenecks that do not impose a severe transition-state penalty. In this sense, fast ion transport depends on the coupling between site topology and site energetics[5,111]. The question is not only where mobile ions can reside, but how easily they can move between nearby configurations and whether those local hops connect into a percolating network over longer length scales.

The framework's topology remains central because it determines the network's dimensionality and resilience. Three-dimensional connectivity is especially advantageous because it allows ions to bypass unfavorable local regions and reduces the vulnerability of long-range diffusion to blocking by grain boundaries, local disorder, or compositional fluctuations. However, topology alone is never sufficient. Even an apparently well-connected framework can remain a poor conductor if the local site energies are too uneven or if bottlenecks are too constricted. This is why materials with similar average structures can exhibit very different ionic conductivities: the decisive difference often lies not in the existence of a geometric channel in the average structure, but in whether that channel corresponds to a statistically connected sequence of low-barrier local migration events.

The nature of the migration bottleneck is equally important because it reflects how the surrounding coordination environment responds during ion motion[112,113]. In oxides, strong metal–oxygen bonding often makes transport highly sensitive to octahedral tilting, cation ordering, and vacancy arrangement, since even subtle geometric changes can significantly alter the bottleneck[114]. In sulfides, larger and more polarizable sulfur anions can stabilize transition states more effectively, allowing local environments to relax more readily during hopping and thereby broadening the low-barrier event landscape[64,65]. Halides show that a framework need not be maximally soft to support rapid transport: in many cases, small local displacements of halide ions are sufficient to connect neighboring shallow minima without requiring extensive framework rearrangement[92,115,116]. In antiperovskites, bottleneck accessibility can be influenced not only by composition and symmetry, but also by dynamic processes such as $OH^-$ reorientation or rotational motion of cluster-like anions[88,105,106]. Across all of these systems, the most favorable conductors are those in which the local coordination environment can accommodate ion motion without sharply destabilizing the transition state.

Defect chemistry adds an additional layer of control by regulating both the density and the connectivity of transport-active local environments. In some materials, defects merely enhance an already viable network;

in others, they are the necessary condition for long-range conduction. Antiperovskites illustrate this most clearly, since their Li-rich topologies are intrinsically well connected but require vacancies, off-stoichiometry, or aliovalent substitution to activate macroscopic diffusion[117]. Yet the same principle applies broadly. In oxides, vacancy concentration and cation disorder often determine whether a rigid framework remains blocked or becomes transport-active[118,119]. In sulfides, local vacancy redistribution and partial site occupancy can flatten the migration-energy landscape[120]. In halides, especially mixed-anion systems, defect chemistry becomes intertwined with local chemical heterogeneity because a second anion modifies charge compensation, polyhedral distortion, and the relative energies of neighboring Li sites simultaneously[69,121]. Productive defect chemistry, therefore, does not simply maximize disorder; rather, it increases the density and connectivity of low-barrier local migration events without destroying the coherence or chemical viability of the host framework.

Anion polarizability and framework deformability contribute in a similarly general way. Their significance lies less in a simple rigid-versus-soft dichotomy than in how they reshape the migration-energy landscape. Sulfides demonstrate that highly polarizable anions can broaden bottlenecks and stabilize transition states across a wide range of local environments. Halides show that comparable transport advantages can also arise from smaller but highly cooperative local displacements within close-packed anion lattices. Mixed-anion halides extend this idea further by redistributing local polarizability and bond stiffness through oxygen, nitrogen, or more complex polyanionic components, thereby generating chemically diverse but still transport-active local environments. In this broader sense, fast ion conduction is promoted whenever the framework reduces the energetic contrast between neighboring migration events and prevents excessive localization of the mobile ion.

This event-connectivity perspective is especially powerful because it bridges crystalline and disordered conductors without forcing them into the same structural language. In highly ordered oxides, the dominant low-barrier events may still correspond closely to crystallographically defined pathways. In sulfides, especially those with substantial site disorder or glass-ceramic character, transport is already better understood as motion through a broader ensemble of accessible local hops. In conventional halides, the coexistence of quasi-degenerate tetrahedral and octahedral environments means that conduction is often better described in terms of a shallow migration landscape rather than a single unique route. In mixed-anion halides and disordered halide-derived systems, this perspective becomes even more important because chemically distinct local coordination environments can all contribute to long-range transport so long as they remain statistically connected. The same logic applies to dynamically assisted antiperovskites, where vacancy topology and anion motion together determine whether local jumps remain continuously accessible over time.

Viewed in this way, the major electrolyte families represent different strategies for constructing and maintaining a connected network of low-barrier migration events. Oxides tend to rely on precise framework geometry and carefully controlled defects to preserve a relatively limited but effective set of transport-active configurations. Sulfides broaden the transport landscape through anion softness and local structural adaptability. Halides achieve efficient transport through small site-energy differences and subtle anion adjustments within structurally coherent close-packed frameworks. Mixed-anion halides demonstrate that local chemical complexity can be used constructively to diversify migration-active environments, while antiperovskites show that a highly connected Li-rich topology can become transport-competitive when defect chemistry and dynamic anion response are tuned appropriately. Across all of these cases, fast ion transport is best understood not as the consequence of one privileged diffusion pathway, but as the macroscopic expression of a connected ensemble of local low-barrier migration processes.

From this standpoint, the most useful design target for future solid electrolytes is not simply the lowest isolated migration barrier or the clearest crystallographic channel. More generally, high performance should be expected when the framework generates a large number of energetically accessible local hops, when these hops remain statistically connected across the structure, and when this transport-active network survives the chemical, electrochemical and mechanical conditions of realistic battery operation. In other words, the challenge is not only to lower individual barriers, but to design solid electrolytes in which low-barrier migration events remain percolatively connected across multiple structural and interfacial length scales. This broader event-connectivity framework provides a more general and transferable basis for understanding fast ion transport across oxide, sulfide, halide and antiperovskite solid electrolytes.

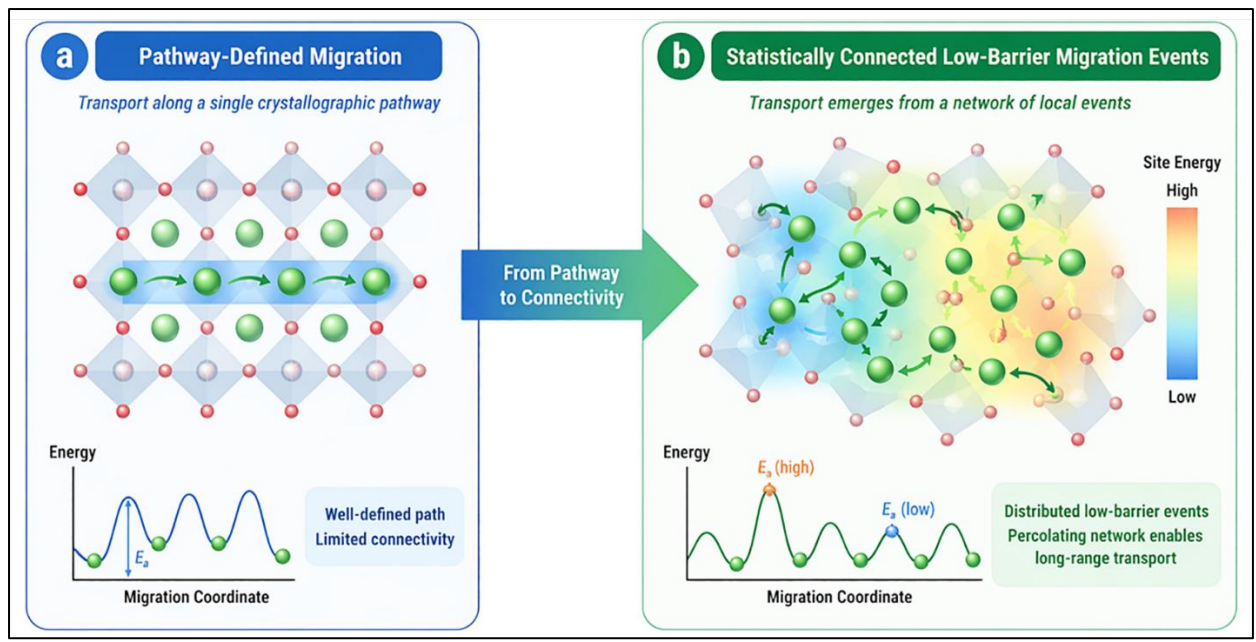

**Figure 5. From pathway-defined migration to statistically connected low-barrier migration events.** (a) Ion transport described as motion along a single crystallographic pathway with well-defined hopping sites. (b) Ion transport emerging from a network of local migration events, where variations in site energies and barriers create a connected landscape of low-barrier pathways enabling long-range diffusion.

## 4. Tools for establishing structure–property relationships

Establishing structure–property relationships in solid electrolytes necessitates more than mere identification of a nominal crystal structure. Ion transport within these materials is concurrently influenced by several factors, including average framework topology, local coordination diversity, defect populations, dynamic disorder, microstructural connectivity, and interfacial evolution. Consequently, no single analytical method can comprehensively capture all of these variables. The most informative investigations, therefore, rely on a combination of experimental and computational techniques that examine different structural levels and timescales, allowing for the understanding of transport not as an isolated material property but as a manifestation of a multiscale structural hierarchy. This multiscale and complementary framework, spanning atomic to macroscopic length scales and relevant timescales and linking structure, local dynamics, and ion transport, is schematically illustrated in Figure 6.

Diffraction-based techniques serve as an essential foundation since they define the average structural framework within which ionic migration occurs. Powder X-ray diffraction and neutron diffraction techniques elucidate phase symmetry, lattice metrics, site occupancies, and the broader geometry of transport-relevant polyhedra[122]. In the context of oxide electrolytes, these methods are critical for distinguishing polymorphs with significantly different transport behaviors, identifying ordering tendencies that hinder conductivity, and assessing the effects of aliovalent substitution on framework symmetry. In halides and antiperovskites, diffraction techniques delineate the close-packed anion topology, the connectivity of metal–halide polyhedra, and the symmetry alterations accompanying transport-relevant structural transitions. Nonetheless, average-structure methods possess considerable limitations; they frequently underrepresent lithium disorder, obscure local fluctuations among various coordination environments, and become increasingly less informative in systems characterized by mixed anions, nanocrystallinity, or amorphous structures, where transport is governed by motifs that only minimally reflect the average unit cell[123-125].

This context underscores the emerging significance of total scattering and pair distribution function analysis within contemporary solid-electrolyte research. These approaches extend beyond traditional crystallography by probing short- and medium-range atomic correlations directly, making them particularly valuable when transport mechanisms depend on local structural diversity rather than strict periodicity[126]. In halides, total scattering can reveal fluctuating tetrahedral and octahedral lithium coordination environments, broad distributions of lithium–chlorine distances, and local deviations from the idealized close-packed configuration. In mixed-anion halides, total scattering techniques can identify chemically distinct cation-centered polyhedra that are otherwise averaged out in conventional refinements[127]. In glass-ceramic sulfides and disordered oxide derivatives, these techniques are among the few means to determine whether local connectivity remains conducive to rapid ion transport, even when long-range order is diminished[128]. Neutron total scattering is particularly critical when resolving lithium, hydrogen, or nitrogen, and is therefore especially effective for antiperovskites, hydroxide-containing phases, and nitrohalides, where light-element coordination and vacancy-related local structures are essential to transport[127,129].

Local spectroscopies provide an additional indispensable layer of insight. Solid-state nuclear magnetic resonance (NMR) is particularly powerful as it establishes a direct link between structure and dynamics: it can resolve multiple local environments, elucidate lithium exchange between them, and identify the temperature range over which motional averaging initiates[130-133]. Thus, NMR is uniquely suited for materials in which the interrelationship between local coordination diversity and transport is tightly intertwined, including disordered sulfides, halides with quasi-degenerate lithium sites, and mixed-anion systems exhibiting broad distributions of local site energies. X-ray absorption spectroscopy and other element-specific local probes are likewise crucial when transport is influenced by mixed-anion coordination or cation distortion, as seen in oxyhalides and nitrohalides. Additionally, vibrational spectroscopies, such as Raman and infrared spectroscopy, can provide valuable information for thiophosphate-based sulfides and dynamically active antiperovskites, as they can track polyanion connectivity, hydroxide reorientation, and borohydride rotation, all of which directly affect the migration landscape.

Transport phenomena must be investigated across multiple timescales. Electrochemical impedance spectroscopy is currently the most prevalent technique for quantifying ionic transport due to its ability, in principle, to distinguish between bulk, grain-boundary, and interfacial contributions[134-137]. This differentiation is essential, as solid electrolytes often exhibit excellent intrinsic conductivity; however, their practical performance can significantly degrade when grain boundaries, pores, or interphases are introduced. The oxide field has established that bulk transport and device-level transport are not synonymous, a lesson that also holds true for sulfides, halides, and antiperovskites. Methods such as nuclear magnetic resonance (NMR) relaxation, line-shape analysis, and exchange measurements offer complementary insights into jump frequencies and local dynamics, often preceding the full establishment of long-range diffusion. Quasielastic neutron scattering further enhances this understanding by probing diffusive motion on timescales and length scales particularly relevant to ion hopping. In the context of antiperovskites and mixed-anion systems, such dynamic techniques are often crucial, as they distinguish between vacancy-mediated hopping and transport processes associated with anion reorientation or dynamic bottleneck modulation.

Computational methods have also become integral to the contemporary analysis of structure–property relationships[138,139]. Density functional theory is widely employed to assess phase stability, compare site energies, evaluate defect formation energetics, and estimate migration barriers through methodologies such as nudged elastic band calculations[17,61,99,140-142]. These computational tools are particularly valuable for identifying feasible transport topologies and rationalizing compositional trends. However, static barrier calculations are frequently inadequate when transport mechanisms are influenced by local disorder, dynamic coordination changes, or correlated motions. Consequently, ab initio molecular dynamics and rigorously developed atomistic simulation approaches are increasingly significant. They enable the observation of how local structural environments evolve over time, how ions circulate among quasi-degenerate sites, how dynamic anion responses stabilize transition states, and whether individual local jumps coalesce into a coherent transport network[56,99]. In halides, such simulations are crucial for capturing fluctuations in tetrahedral–octahedral lithium coordination and subtle adjustments in halide ion positions[143,144]. In antiperovskites, they elucidate the interaction between vacancy movement and anion dynamics[117]. In disordered mixed-anion systems, these simulations provide a clear methodology for determining whether transport is governed by a single dominant pathway or a network of connected low-barrier events[127].

Operando and *in situ* methodologies have gained prominence, as the structures governing transport are often not static. Mechanochemical synthesis, thermal treatment, cycling-induced interphase formation, and atmospheric exposure can reshape the transport-active network. *In situ* diffraction during milling or heating can monitor the extent to which crystalline precursors reorganize, disorder, or partially amorphize into transport-active states[123]. Operando impedance, spectroscopy, and microscopy can verify whether the conductive network identified in isolated pellets remains intact when the electrolyte interacts with electrodes or undergoes electrochemical cycling[124]. This verification is particularly crucial for sulfides and halides, where intrinsic transport is frequently entangled with interphase formation and chemical instability, and for antiperovskites, whose symmetry, defect structure, and anion dynamics can vary significantly with temperature.

Cumulatively, these methodologies highlight a broader methodological insight. Structure–property relationships in solid electrolytes must not be approached as a series of disconnected measurements but rather as a coordinated effort to elucidate a hierarchy of transport-relevant variables. The fundamental inquiries pertain not only to the average framework but also to how local environments diverge from that average, how defects are distributed, which migration events are dynamically accessible, whether those events are spatially and temporally interconnected, and how processing or interfaces may modify that connectivity. The most enlightening studies are those that integrate crystallography, local structural analysis, dynamical measurements, and atomistic simulations into a cohesive, comprehensive framework.

This integrated approach is particularly significant for halides and their derivatives. In conventional halides, average diffraction techniques can effectively define the close-packed anion topology. However, to accurately elucidate the quasi-degenerate lithium environments and the subtle halide adjustments that underpin transport phenomena, it is essential to employ total scattering, nuclear magnetic resonance (NMR), and computational simulations. In the case of mixed-anion halides, the utilization of local structural probes becomes even more critical, as average structures frequently conceal the chemically distinct coordination motifs that influence site energies and create bottlenecks. For antiperovskites, traditional diffraction methods establish the lithium-rich topology; however, a comprehensive understanding of how vacancies, symmetry regulation, and anion motion activate this topology requires combining defect analysis, transport measurements, and dynamical probing techniques. Consequently, the future of structure-property analysis in solid electrolytes will not rest on the identification of a singular optimal tool, but rather on the development of methodological workflows that are capable of resolving transport phenomena as an interconnected hierarchy of structural events.

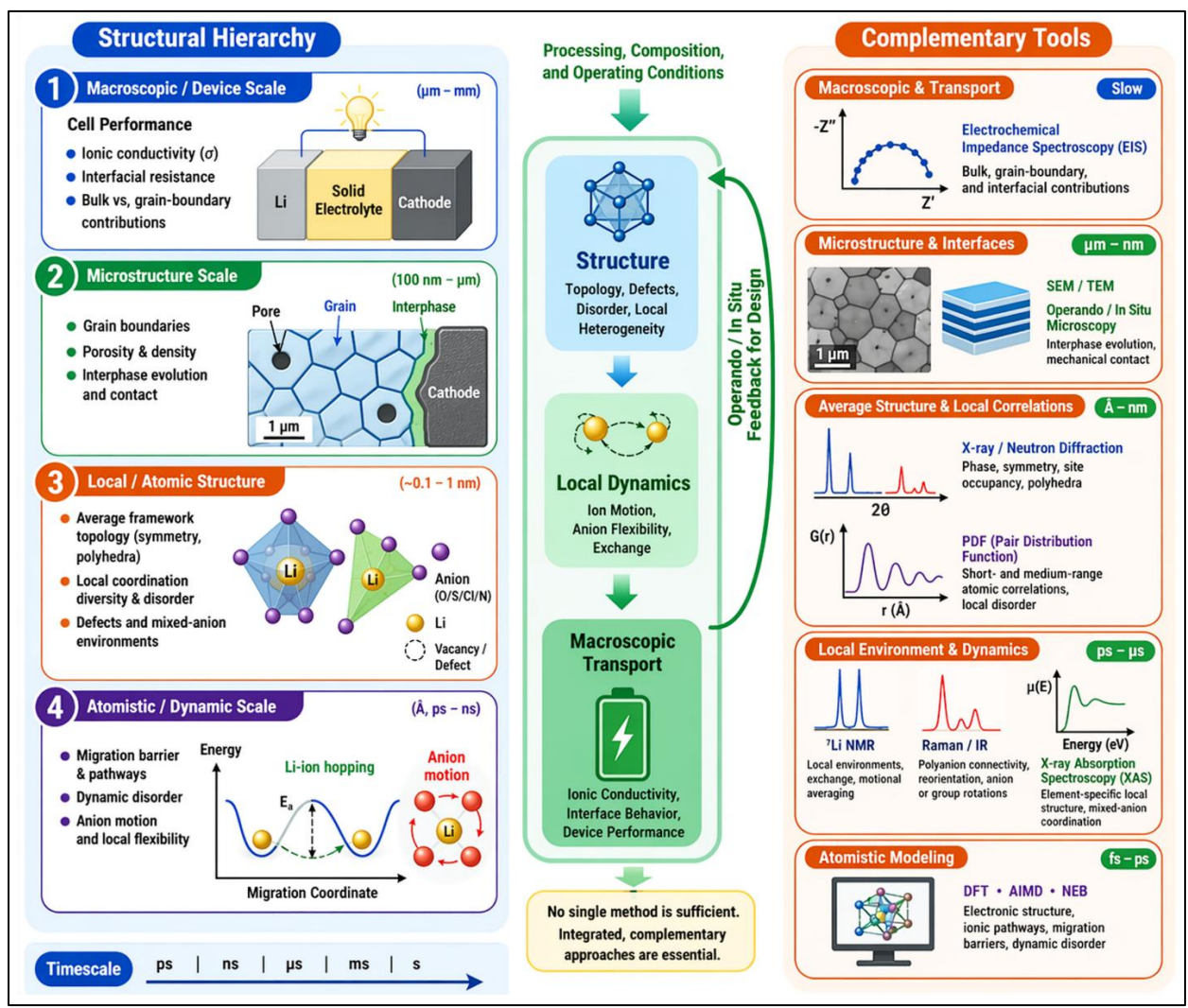

**Figure 6. Multiscale and complementary tools for establishing structure–property relationships in solid electrolytes.** Techniques spanning atomic to macroscopic length scales and relevant timescales are integrated to connect structure, local dynamics, and ion transport.

## 5. Outlook

The forthcoming phase of research on solid-state electrolytes is unlikely to be characterized solely by the ongoing pursuit of enhanced ionic conductivity at room temperature. For oxide, sulfide, and halide systems alike, the primary challenge has shifted from identifying fast-ion conductors in isolation to developing comprehensive design strategies that integrate transport-active structures with electrochemical stability, interfacial compatibility, and feasible manufacturability. In this context, a significant shift in scientific understanding is required: solid electrolytes should increasingly be regarded not merely as materials with a singular intrinsic conductivity, but as dynamically evolving ion-transport networks whose performance depends on the maintenance of transport-relevant local environments across multiple length scales and under realistic operational conditions.

From this perspective, a critical avenue for future research involves the transition from structure identification to the engineering of transport landscapes. Considerable progress has already been made in identifying the types of framework associated with rapid ion conduction. However, the next challenge is more nuanced: it involves determining how to appropriately tune local coordination diversity, defect populations, anion mixing, and dynamic structural responses to effectively reshape the migration-energy landscape in a controlled and predictive manner. This is particularly pertinent for halide-based systems, where ion transport typically arises not from a single dominant crystallographic pathway but from the coexistence of various weakly differentiated local environments. In such materials, the essential design consideration is no longer solely focused on widening bottlenecks or creating vacancies; rather, it is imperative to preserve the statistical connectivity of low-barrier migration events while averting deep energetic traps or unstable local chemistries.

On the other hand, a significant area of current research concerns the intentional incorporation of local chemical complexity. Systems derived from mixed anions and halides have shown that local coordination heterogeneity does not necessarily impede ion transport; rather, it can expand the transport landscape and create novel migration-active environments under appropriate conditions. Consequently, the central inquiry for forthcoming research should not be whether disorder should be integrated, but rather what type of disorder may be conducive to enhancing transport. This distinction is critical. Productive local heterogeneity ought to minimize site-energy disparities, maintain short-range connectivity, and eliminate electronically or chemically unstable motifs. Conversely, disorder that obstructs percolation or creates environments that are significantly trapping will remain detrimental, even if it appears structurally analogous on average. Thus, a major opportunity exists in formulating predictive frameworks to determine when mixed-anion complexity, local distortion, and partial amorphization may improve transport and when these factors may destabilize the conduction network.

This issue is intrinsically linked to the future of halide solid electrolytes and their derivatives. The rapid advancement of conventional halides has demonstrated the benefits of closely packed anion frameworks with quasi-degenerate lithium environments. However, the more transformative potential may reside in transcending simple chloride- and bromide-based lattices towards more versatile halide-derived coordination networks. Oxyhalides, nitrohalides, sulfate-containing halide derivatives, and materials related to antiperovskites indicate that halide chemistry should be perceived not merely as a singular materials class but as a platform for constructing transport-active frameworks characterized by tunable local symmetry, defect chemistry, and anion responsiveness. The challenge moving forward is to translate this flexibility into design principles that are sufficiently general to assist in material discovery. Specifically, future investigations should elucidate how the incorporation of mixed anions influences not only conductivity but also hydrolysis pathways, reductive stability, and the chemistry of interphase formation, as these properties are likely to be interrelated in practice.

Another significant frontier involves redefining the description of transport phenomena in increasingly complex solid electrolytes. In highly ordered crystalline materials, the established language of crystallographic diffusion pathways remains appropriate and effective. Nevertheless, this framework becomes progressively inadequate as local disorder, dynamic anion mobility, and coordination diversity increase. A more comprehensive and potentially more potent approach is to characterize long-range ion conduction as the macroscopic consequence of statistically connected low-energy migration events distributed throughout the local structure. This paradigm shift carries substantial implications, suggesting that the pertinent descriptor for future solid electrolytes may be the density, distribution, and connectivity of transport-active local environments rather than merely the lowest isolated migration barrier or the existence of a singular idealized channel. Developing metrics that effectively capture this event-connectivity landscape may emerge as one of the most critical theoretical and computational endeavors in the field.

The anticipated shift in solid electrolyte research will necessitate methodological advancements. The critical structure–property relationships in contemporary solid electrolytes increasingly lie at the interface between average order and local heterogeneity, a domain that conventional crystallography is often inadequate to resolve. Future progress will be contingent upon a more integrated approach that combines diffraction, total scattering, local spectroscopy, atomistic simulation, and operando characterization, not merely as parallel methodologies but as integral components of a unified framework. This framework aims to elucidate how transport-active networks are formed, evolve, and degrade over time. This requirement is particularly pressing for halide-derived and antiperovskite systems, where it is essential to concurrently consider quasi-degenerate lithium coordination, mixed-anion local structure, vacancy topology, and dynamic anion-assisted hopping.

Moreover, there exists a critical challenge in transcending equilibrium structure–property relationships to address the reality that many of the most pertinent solid electrolytes are metastable, processing-dependent, or subject to dynamic evolution. Processes such as mechanochemical synthesis, pressure-assisted densification, local decomposition, and interphase growth can yield structures that are not represented by conventional equilibrium phase descriptions, yet they exert substantial influence on transport characteristics. This is especially relevant for sulfides and halide-derived materials, in which local disorder or mixed-anion coordination may facilitate transport while also conferring chemical fragility. Consequently, future design strategies must incorporate not only the structure of the synthesized electrolyte but also its developmental trajectory during fabrication and operation. In this regard, the inquiry extends beyond identifying which frameworks are conducive to considering which transport-active networks can maintain connectivity, exhibit chemical self-limitation, and ensure mechanical viability throughout the operational lifespan of a practical cell.

Ultimately, a significant advancement may lie in a shift in design philosophy. Rather than regarding conductivity, stability, and processability as distinct optimization targets, the future of solid-electrolyte research must increasingly perceive them as co-emergent outcomes of framework chemistry and local structural organization. This approach may be particularly promising for halides and their derivatives, which already encompass a diverse range of structural regimes, from closely packed crystalline lattices to mixed-anion disordered frameworks and lithium-rich antiperovskite topologies. Thus, these materials present a unique opportunity to explore whether transport-active local complexity can be rendered not only rapid but also chemically and mechanically dependable.

In this broader context, the future of solid-state electrolyte design may hinge less on identifying a singular dominant materials family and more on understanding how various framework chemistries encode transport-active connectivity in diverse manners. Oxides, sulfides, and halide-derived systems should not merely be regarded as competing classes of electrolytes but as complementary structural models that offer distinct solutions to the same fundamental challenge: establishing and maintaining a connected network of low-barrier ionic migration events under realistic electrochemical conditions. The central objective moving forward is to convert this qualitative insight into a predictive design framework. Only through this transition will the field progress from merely discovering fast ion conductors to engineering solid electrolytes with the requisite reliability, tunability, and functional integration necessary for practical all-solid-state batteries.

## Acknowledgements

This work was supported by Shenzhen Science and Technology Program (KQTD20200820113047086), Shenzhen Key Laboratory of Solid State Batteries (SYSPG202412111173726011), and Guangdong Provincial Key Laboratory of Energy Materials for Electric Power (2018B030322001), and we also acknowledge the Major Science and Technology Infrastructure Project of Material Genome Big-science Facilities Platform.